\documentclass[12pt]{article}

\usepackage{amsmath, amsfonts, amssymb, latexsym, amsthm, thmtools}
\usepackage{bbm} 
\usepackage[letterpaper, margin=1in]{geometry}
\usepackage[T1]{fontenc}
\usepackage[utf8x]{inputenc}
\usepackage{lmodern}
\usepackage{authblk}
\usepackage{graphics}
\usepackage[english]{babel}
\usepackage{csquotes}
\usepackage{capt-of}
\usepackage{natbib}
\usepackage{setspace}
\usepackage{hyperref}

\usepackage[normalem]{ulem} 
\usepackage[obeyDraft]{todonotes}
\presetkeys{todonotes}{backgroundcolor=orange!40, size=\tiny}{}

\usepackage{algorithm,algorithmic} 

\newcommand{\bP}{\mathbb{P}} 

\newcounter{mytodo}

\newcounter{mycomment}

\begin{document}
\title{A Stochastic Time Series Model for Predicting Financial Trends using NLP}
\author[1]{Pratyush Muthukumar}
\author[2]{Jie Zhong}
\affil[1]{Department of Computer Science, University of California, Irvine}
\affil[2]{Department of Mathematics, California State University, Los Angeles}

\maketitle

\abstract{Stock price forecasting is a highly complex and vitally important
  field of research. Recent advancements in deep neural network technology allow
  researchers to develop highly accurate models to predict financial trends. We
  propose a novel deep learning model called ST-GAN, or Stochastic Time-series
  Generative Adversarial Network, that analyzes both financial news texts and
  financial numerical data to predict stock trends. We utilize cutting-edge
  technology like the Generative Adversarial Network (GAN) to learn the
  correlations among textual and numerical data over time. We develop a new
  method of training a time-series GAN directly using the learned
  representations of Naive Bayes' sentiment analysis on financial text data
  alongside technical indicators from numerical data. Our experimental results show significant improvement over various existing models and prior research on deep neural networks for stock price forecasting.}

\section{Introduction}
Stock market prediction has been a topic of interest among researchers,
corporations, and market enthusiasts for decades. While the lucrative and prophetic
nature of accurate stock market prediction is widely sought after, it may be
difficult to grasp the efficacy of financial forecasting given the volatile
nature of global stock markets.
The latter sentiment is not new; in \citet{cowles}, the US economist
Alfred Cowles wrote that even the most successful stock market forecasters did
``little, if any, better than what might be expected to result from pure
chance''. However, with the rise of big data systems and machine learning, we find that the perceived quality of stochasticity in financial markets may actually contain a pattern embedded in data that spans across financial sectors and fields of interest \citep{lobato1998real}. 

Most machine learning approaches to predicting financial trends focus on an
analysis of numerical features and technical indicators of stocks. Various
machine learning models including Linear Regression, Logistic Regression,
Support Vector Machines (SVMs), Decision Trees, and Random Forests (RFs) have
been implemented to tackle this problem; see the survey paper
\citep{kimoto1990stock} and references therein. More recently, deep learning
models including Artificial Neural Networks (ANNs), Recurrent Neural Networks
(RNNs), and Convolutional Neural Networks (CNNs) have also been used to predict
stock prices  \citep{tsantekidis2017forecasting,sokolov2020neural}. The large
majority of these works utilize machine learning models trained on solely
financial numerical data. However, stock market prediction is an invaluable
market research strategy --- stock prices are determined on the behavior of
retail investors who themselves base decisions on available news sources and
statistical measures. Thus, 
an entirely different category of feature information can be learned through an analysis of textual data. 

The impact of financial news and text data on the trends of the stock market
cannot be understated. The financial market is highly dependent on the decisions
retail investors take, which in turn, are dependent on the daily news and
information sources they read. Thus, by incorporating the information of textual
data to a financial forecasting model in addition to the technical indicators
available, a machine learning model can understand full-fledged patterns of financial trends.

There has been previous work on utilizing Earnings Conference Calls (ECCs),
quarterly public briefings by a company's upper management team, investment
team, and legal team. ECCs provide valuable insight into the financial health of
a company. Recent advancements in the field of Natural Language Processing (NLP)
allow state-of-the-art language models including Google BERT
\citep{devlin-etal-2019-bert} and OpenAI's GPT \citep{radford2019language} series to be applied to financial text data to learn word embeddings and representations which will later be used to predict financial trends. These models rely on discerning the sentiment of words or sentences of a financial textual article, which is referred to as sentiment analysis in the field of NLP. Applying pretrained language models to perform sentiment analysis is just one of the many machine learning models that could be applied to solve this task; others include Naive Bayes classification models, Convolutional Neural Networks, or Recurrent Neural Networks; see the survey paper \citep{das2019zero} and references therein. 

Although we see great strides of innovation in machine learning models which seek to predict financial trends through solely numerical features or solely textual features as input, we seldom see a model that encompasses information of both textual and numerical format as the basis for predicting the stock market. Hence, we seek to utilize advanced machine learning models that extract details from financial text sources and numerical indicators to predict stock trends. We define our domain of interest to be the financial market trends in the aerospace industry. The aerospace industry serves as the perfect litmus test for the rest of the financial market, as the key public and private corporations in this sector contributed to \$151 billion in export sales to the U.S. economy in 2018 alone. The aerospace industry is also inherently global, and according to \citet{dussauge1995determinants}, forecasting the outlook of this industry gives us a glimpse into the world economy as well as the localized United States economy. 

The challenge we faced when constructing such a hybrid model predominantly stems
from the question of effectively joining numerical and textual features (or
rather utilizing the output of a model trained on solely textual features as an
input for the one trained with numerical features). Additionally, it is imperative that we construct a neural network architecture with the capability to learn from data that may be sparse or less correlated, given the nature of financial forecasting.  

We propose a novel deep learning model in this paper which utilizes state-of-the-art machine learning models including the Generative Adversarial Network (GAN) to predict stock trends using financial text data and financial numerical data. We invent a new method of directly using sentiment analysis results on financial text data as an input to the time-series GAN. Our model outperforms existing researched models in deep learning for financial forecasting in various forecasting time horizons. 

The remainder of the paper is structured as follows. Section \ref{sec2} illustrates our key contributions. Section \ref{nb} describes our model architecture. Section \ref{3one} and \ref{3two} outline our experimental testing methodologies. Section \ref{3three} describes our model's experimental results. Section \ref{four} concludes our findings and discusses our future work. 

\section{A Stochastic Time Series Model}\label{sec2}

Our proposed model provides the following technical contributions to the field of financial forecasting with machine learning. 

\paragraph{Robust Textual Understanding of Financial Data}

We utilized state-of-the-art sentiment analysis techniques from the field of NLP trained on continually updated data with high temporal resolution. Moreover, we surpassed the standard depth of use for these sentiment analysis techniques.

\paragraph{Advanced Time-Series Generative Prediction} 

We modified the increasingly popular Generative Adversarial Network (GAN) to predict for data indexed over time through a generative process. \citet{adel2018discovering} has shown that generative models outperform traditional discriminative models like Logistic Regression, Support Vector Regression, and Conditional Random Fields in certain time-series tasks. 

\paragraph{Novel Sequential Textual Embedding Technique} 

The novel aspect of our research is due to the hybridization of the modular
financial textual network and financial numerical network. To the best of our
knowledge, our technique is the first to directly include textual embeddings to
generate stock forecasts through our deep generative model. We invented a new
method to use the financial text sentiment analysis results as a latent space
input to the generator of the Generative Adversarial Network trained on
numerical features. Our results show a $32.2\%$ decrease in averaged NRMSE error
over multiple forecasting time horizons compared to the currently best performing research utilizing other deep learning models for stock price prediction. 

\subsection{Model Architecture} \label{nb}
The structure of our two-stage model follows a sequential organization, where the first unit of our model feeds as input to the second unit. The first section of the model is a Naive Bayes' classifier model to perform sentiment analysis on our textual data inputs.

The goal of sentiment analysis is to systematically identify trends in textual data, often word-by-word or sentence-by-sentence \citep{narayanan2009sentiment}. We performed both on a number of subsets of our textual data. The output of this Naive Bayes' classifier model is a set of sentiments, which is a sentiment from the set \{positive, neutral, negative\}. It is evident why a classification model works best for sentiment analysis, as the classifier predicts the sentiment class for each sentence of a financial textual document. 

The Naive Bayes' algorithm is a fundamental generative model used widely in
machine learning. \citet{tan2009adapting} first proposed using Naive Bayes'
classification for sentiment analysis due to the relative independence of words
in a document. The goal of Naive Bayes sentiment analysis is to use initial
training sentiments of a set of individual words to predict the sentiments of
unknown words and thus the sentiment of each sentence in the document. Specifically, given a set of word vectors $X = \{x_1,x_2,\dots, x_k\}$, we predict
$Y$, which is a label with $M$ possible classes. Models that seek to
find $\mathbb{P}(Y = y | X = \{x_1,x_2,\dots,x_k\})$ directly, are
discriminative. However, as a generative model, Naive Bayes' classifier reverses
the conditioning, 
\begin{equation*}
\mathbb{P}(Y|X) = \frac{\mathbb{P}(X|Y)\mathbb{P}(Y)}{\mathbb{P}(X)},
\end{equation*}
by the classical Bayes formula.

Now, the most likely class of all possible classes $Y$ given the data $X$ is 
\begin{align*}
    \arg\max_{y \in Y} \mathbb{P}(Y|X) &= \arg\max_{y \in Y} \frac{\mathbb{P}(X|Y)\mathbb{P}(Y)}{\mathbb{P}(X)} \\
    &= \arg\max_{y \in Y} \mathbb{P}(X|Y)\mathbb{P}(Y) \\
    &= \arg\max_{y \in Y} \mathbb{P}(x_1,x_2,\dots,x_k|Y)\mathbb{P}(Y).
\end{align*}
In practice, we find that calculating the probability
$\mathbb{P}(x_1,x_2,\dots,x_k|Y)$ is computationally infeasible, as we would
have required $M^k$ computations to find $\mathbb{P}(X|Y)$, where $k$ is the number of words in the document. Instead, we assume independence, i.e., 
\begin{equation*}
    \mathbb{P}(x_1, x_2, \dots, x_k | Y) = \mathbb{P}(x_1|Y) \cdot \mathbb{P}(x_2|Y) \cdots  \mathbb{P}(x_k|Y).
\end{equation*}
Then, the most likely class becomes \begin{equation} \label{eq:1}
    \arg\max_{y \in Y} \mathbb{P}(x_1, x_2, \dots, x_k | Y)\bP(Y) = \arg\max_{y \in Y} \mathbb{P}(Y)\prod_{x_k\in X} \mathbb{P}(x_k|Y).
\end{equation}
The independence assumption here is plausible due to the nature of language:
individual words are not dependent on other words except for context and grammar
\citep{lewis1998naive}. Using independence, we perform the algorithm in linear
time: to compute the most likely class, we perform $M \cdot k$ computations. It
is important to note that the output of Naive Bayes' sentiment analysis on a
document is an $n$-dimensional vector $v$, where each component $v_i \in \{-1,0,1\}$ and $n$ denotes the number of sentences in the document, assuming we are performing sentence-by-sentence Naive Bayes' sentiment analysis. We feed this output vector $v$ as input to a second section of our model in a novel procedure.

The second section in our model is a deep predictive model capable of utilizing the numerical data and textual sentiment analysis embeddings to identify patterns among the input data. To achieve this goal, we select the Generative Adversarial Network (GAN) architecture. The GAN unit was developed as a generative variant of the traditional Convolutional Neural Network used for image-based deep learning \citep{goodfellow2014generative}. GANs generate samples similar to the training data through adversarial training. 

The generative nature of the network allows GANs to be particularly effective for financial data with multiple dimensions, low correlation, or sparsity \citep{muthukumar2020astochastic}. This is because GANs are inherently an unsupervised learning model, since the predictive process of the network does not require ground truth training data. Instead, the generator module of a GAN uses randomly sampled input variables to create a prediction. In financial forecasting, the generative model introduced in \citet{krishnan2018challenges} provides benefits which allow for higher quality prediction on high dimensional, sparse data.

 The Generative Adversarial Network is comprised of two sub-networks: the \emph{generator} and the \emph{discriminator}. The generator is a neural network which traditionally uses a fixed-length random vector as input to generate a sample. The discriminator is another neural network which traditionally classifies between training data and generated samples. It is important to note that the generator is never exposed to the real world training data, as that is only available to the discriminator. The generator and discriminator engage in a two-player minimax game throughout the process of training, where the discriminator seeks to minimize the discriminator loss or accurately classify between a real and generated sample, while the generator seeks to maximize the discriminator loss or ``fool'' the discriminator. 

Let the generative network $G$ have a latent space input $x_z \sim p_z,$ where $x_z$ acts as a random ``seed'' for the generator to sample. In a traditional GAN, $x_z$ is a set of latent variables sampled from a prior random noise distribution $p_z$ \citep{voynov2020unsupervised}. Let the discriminative network $D$ have input either $x_t \sim p_t$ or $x_g \sim p_g$ where $x_t$ and $x_g$ denote the ground truth training data and generated samples respectively. 

The loss function for the GAN, described in \citet{goodfellow2014generative}, is derived from the binary cross entropy loss and is defined as
\begin{equation*}
    E(G,D)= \frac{1}{2}(\mathbb{E}_{x\sim p_t}[1-D(x)] + \mathbb{E}_{x\sim p_g}[D(x)]),
\end{equation*}
where the goal of the algorithm is \begin{equation*}
    \max_G \bigg( \min_D \bigg(E(G,D)\bigg) \bigg) = \max_G \bigg( \min_D \bigg(\frac{1}{2}\big( \mathbb{E}_{x\sim p_t}\big[1-D(x)\big] + \mathbb{E}_{x\sim p_g}\big[D(x)\big] \big) \bigg) \bigg).
\end{equation*} Intuitively, the discriminator seeks to minimize the likelihood of the discriminator incorrectly classifying samples, while the generator seeks to maximize this likelihood $E(G,D)$. 

An oversight of the traditional GAN model is that the input of the generator is
sampled from a prior noise distribution without information from the training
dataset. Recently, \citet{chen2016infogan} proposed the model InfoGAN, which
alters the minimax loss function to include characteristics of the random latent
space input $x_z$. This allows the generator and discriminator to iteratively
improve the latent space variables, but the initialization of the latent space
variables remains as data points from a random sample. 

We present a novel method of inputting
relevant high-level embeddings as a replacement to the random latent inputs
$x_z$ used by the Generator model of the GAN. The key idea of our work lies in the way in which the Naive Bayes' sentiment
analysis outputs are used as inputs for the Generative Adversarial Network. Our
architecture allows us to use high-level embeddings of financial text data as an
additional input when predicting for future stock trends on a deep neural
network with financial numerical data inputs. The Naive Bayes' sentiment
analysis on financial news texts produces an $n$-dimensional vector $v$, where
$n$ denotes the number of sentences in the text data document. We use this
vector $v$ as the latent variable ``seed'' $x_z$ in the GAN model, which uses
financial numerical data as the ground truth training data to generate
predictions. We build on the work from InfoGAN by not only using the tweaked
discriminator loss function that includes the latent variable $x_z$ as a
tuneable parameter, but also directly initializing $x_z$ with the
$n$-dimensional vector $v$, instead of sampling from random noise distribution
$p_z$.

In the traditional GAN model supported in the PyTorch module, the default value
for the latent vector $x_z$ is a 100 dimensional vector created by sampling 100
points from a standard normal distribution. For our model, instead, we defined $x_z$ in our PyTorch GAN model using the Naive Bayes' Sentiment Analysis vector output. To do so, we select the top 100 sentiment analysis outputs that had the highest confidence probability. That is, the top 100 highest probabilities of the most likely class of all financial text documents analyzed, defined as 
\begin{equation*}
    \arg\max_{y \in Y} \mathbb{P}(Y)\prod_{x_k\in X} \mathbb{P}(x_k|Y),
\end{equation*}
are selected to form the 100 dimensional vector $x_z$. We then normalize the resulting vector such that the mean of the points is zero and the standard deviation is one. That is, for each $x_z^{(i)} \in x_z,$
\begin{equation*}
    x_z^{(i)} := \frac{x_z^{(i)} - \mu_{x_z}}{\sigma_{x_z}},
\end{equation*}
 where $i = 1,2,\dots, 100$, $\mu_{x_z}$ is the mean of the elements in the vector $x_z$, and $\sigma_{x_z}$ is the standard deviation of the elements in $x_z$.

 Thus, we have initialized our latent variables for the generator network of the GAN model with the scaled vector of the top 100 strongest sentiments of sentences in financial news documents for a stock. The GAN model continually updates the initialized latent vector $x_z$ through the tweaked discriminator network loss function 
 \begin{equation*}
    E(G,D)= \frac{1}{2}(\mathbb{E}_{x\sim p_t}[1-D(x)] + \mathbb{E}_{x\sim p_z}[D(x)]), 
\end{equation*} 
as described in \citet{chen2016infogan}.

In this way, we are able to ensure that meaningful representations from the
output of Naive Bayes' sentiment analysis on financial texts are used as an input to a Generative Adversarial Network trained on financial numerical data. We also ensure that throughout the training process, the latent variable input is fine-tuned for better prediction accuracies. 

\section{Experiments}

\subsection{Data Analysis} \label{3one}

\begin{figure}
   \centering
    \includegraphics[width=\textwidth]{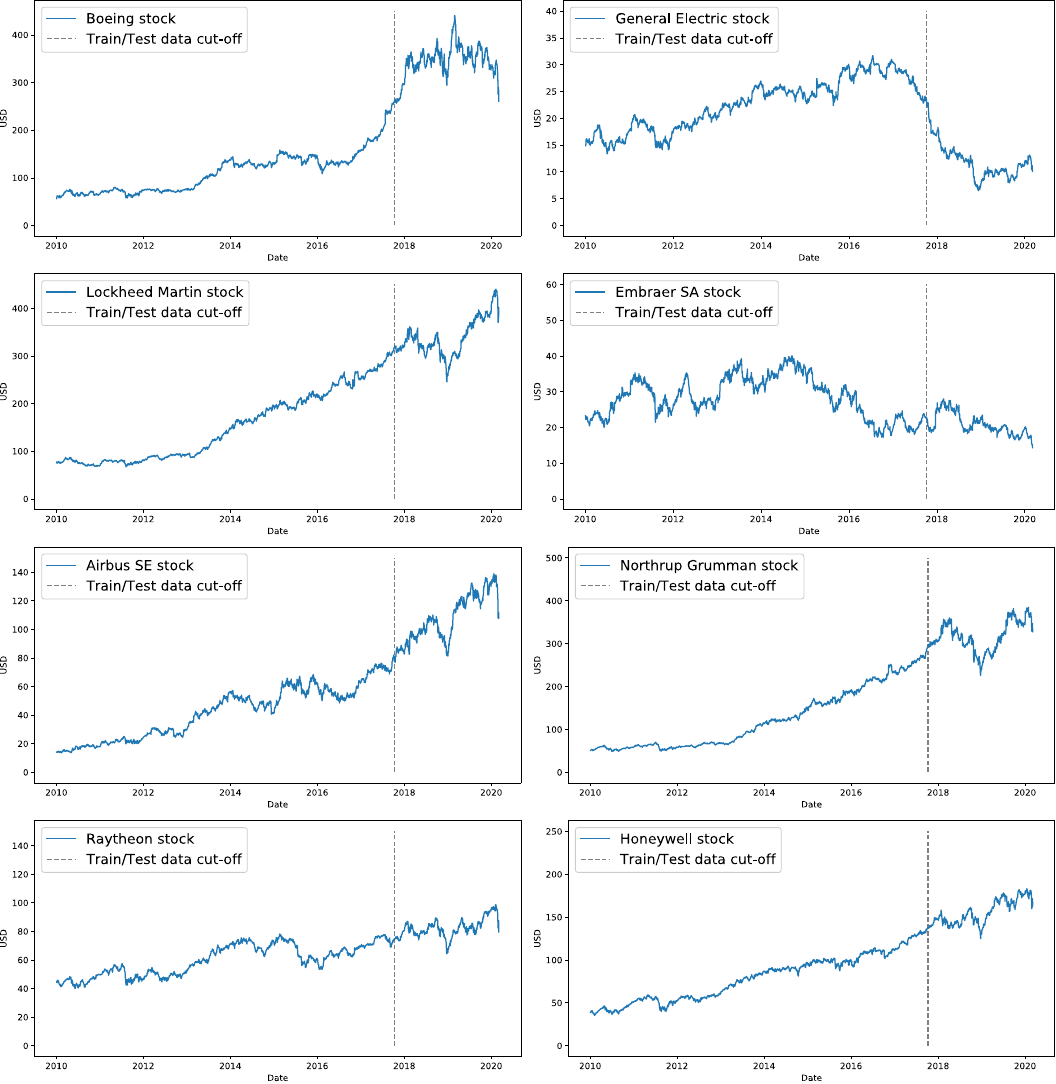}
    \caption{Visualization of Daily Historical Close Prices of 8 Aerospace Stocks}
    \label{fig:histData}
\end{figure}

Our numerical dataset of financial indicators was sourced from the Yahoo Finance historical stock price dataset. We select 8 stocks to apply in our model, all of which represent a company in the aerospace industry.  We focus on the financial trends of the aerospace industry because this microcosm of the financial sector accurately responds to the long-term and short-term effects of current events globally \citep{bae1996empirical}. The eight stocks we select are BA (Boeing), LMT (Lockheed Martin), AIR.PA (AirBus SE), RTX (Raytheon), GE (General Electric), ERJ (Embraer SA), NOC (Northrup Grumman), and HON (Honeywell). This portfolio of aerospace stocks is highly diverse and global, with corporation headquarters in North America, South America, and Europe. We use historical daily stock prices for all eight aerospace stocks from January 1, 2010 to March 6, 2020. We choose this timeframe as a decade worth of data exhibits both short-term and long-term effects of current events on the financial sector. The training data spans from January 1, 2010 to January 24, 2020. The testing data spans from January 25, 2020 to March 6, 2020. The stock data includes the maximum price, minimum price, opening price, closing price, adjusted closing price, and volume for each day. Figure \ref{fig:histData} displays a visualization of the daily closing prices in USD for the eight stocks throughout the timeframe selected. 

\begin{figure}
    \centering
    \includegraphics[width=\textwidth]{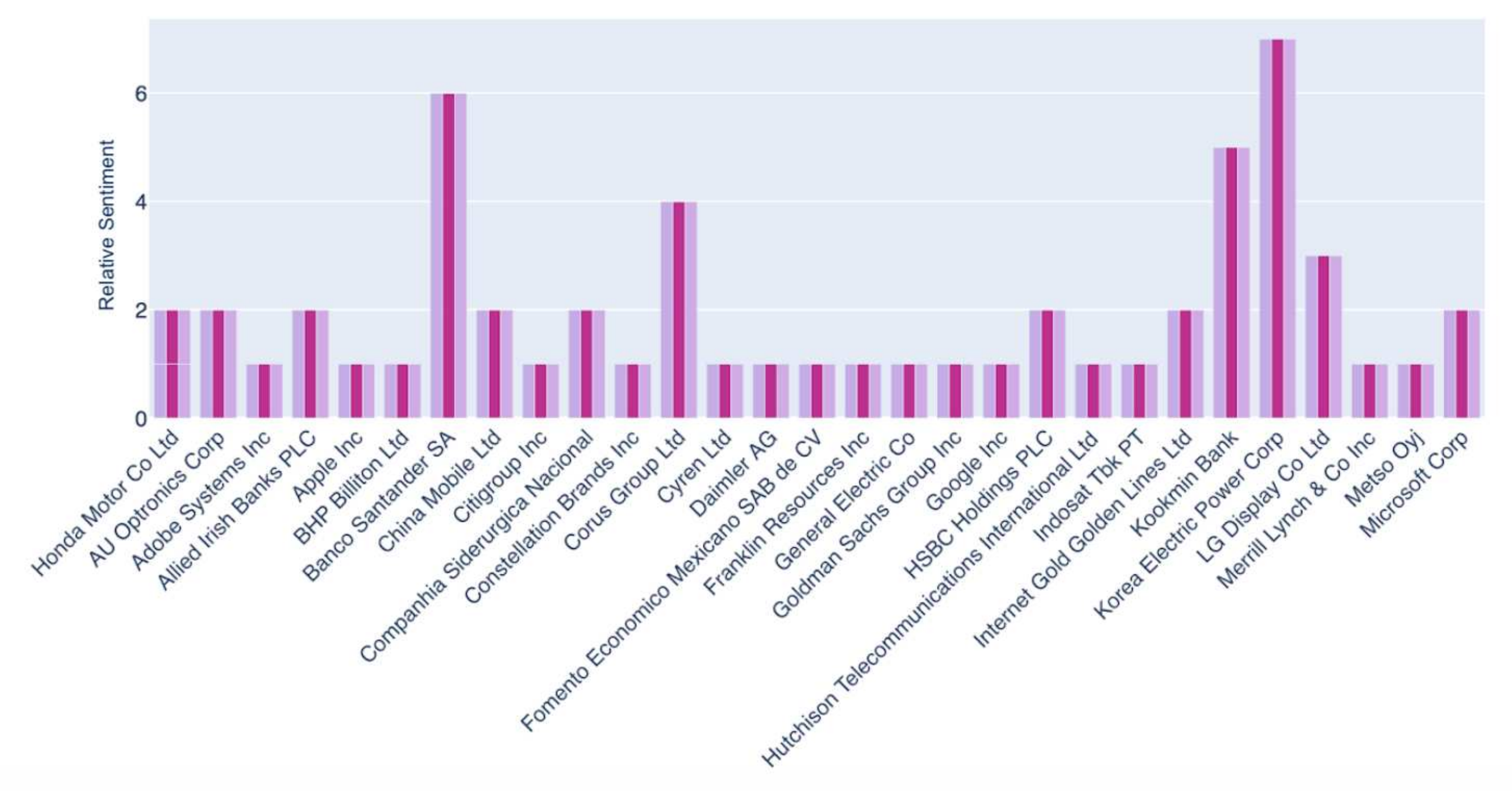}
    \caption{Non-aerospace Company Name Relative Sentiment Results for Context-Based Sentiment Analysis on Boeing Financial News Documents}
    \label{fig:company_names}
\end{figure}

Our financial text dataset is collected from various financial news outlets. We primarily use financial texts from Seeking Alpha, while also using texts from Forbes, MarketWatch, and Twitter. We created a webscraper to collect and download news articles where the headline contained at least one of the eight names of the aerospace stocks we selected. The bulk of our financial texts are in the form of Earnings Conference Calls (ECCs). The subject and keywords spoken during these ECCs can have direct impacts on the future stock trends of a company, so we analyze the sentiments of the ECCs using the Naive Bayes' classifier. 

For the financial text documents, in addition to applying the Naive Bayes' classifier on each sentence to generate the latent vector input to the generator network of the GAN model, we also perform other variants of sentiment analysis, statistical calculations, and pre-processing techniques to analyze the financial texts further. 
We invent and perform context-based sentiment analysis on company names mentioned in the text documents that aren't the eight aerospace companies used we are predicting. Here, context-based sentiment analysis is the technique in which we average the sentiment values of words found around a specific keyword to determine its relative sentiment. 

\begin{figure}
    \centering
    \includegraphics[width=\textwidth]{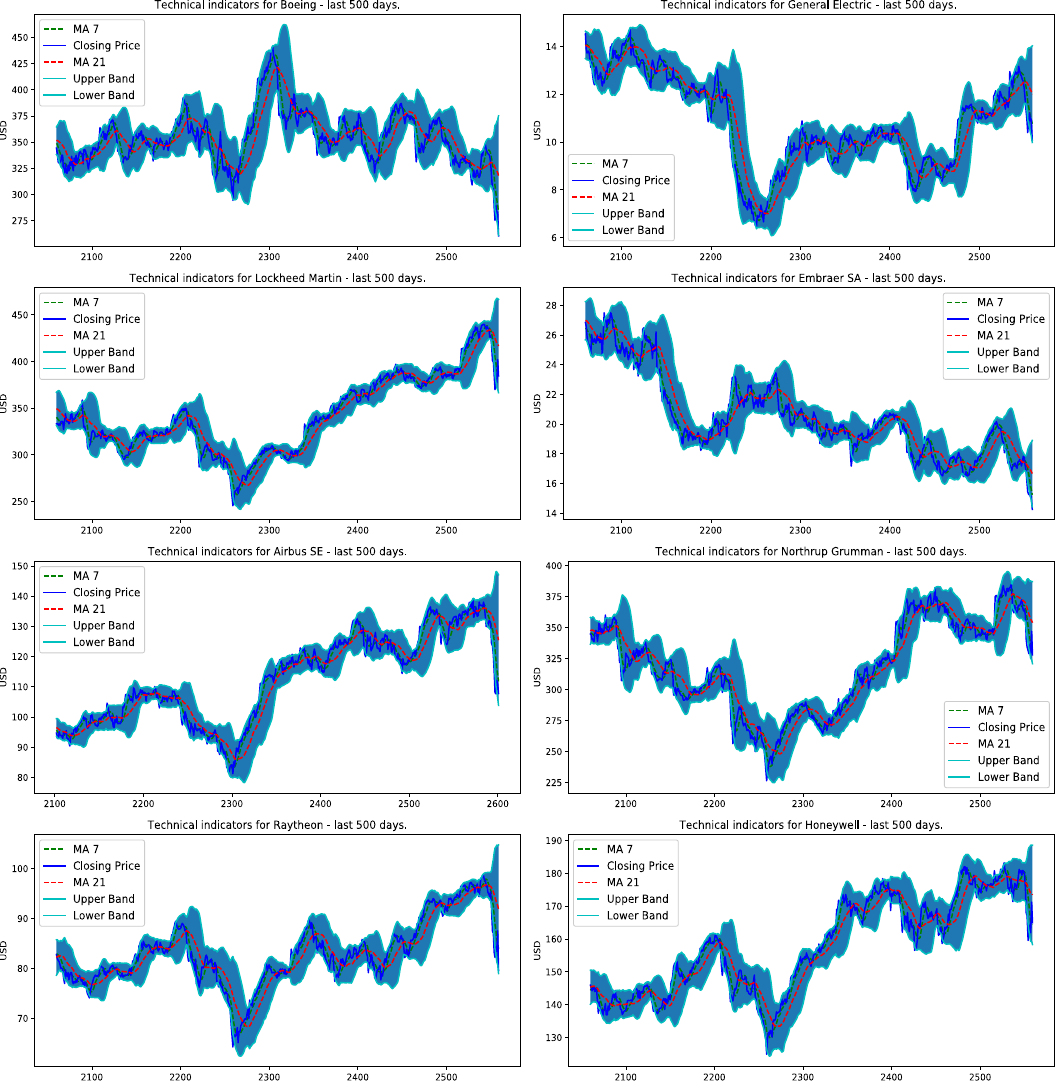}
    \caption{Visualization of a Subset of the Financial Indicators Calculated from the Raw Close Prices for each of the 8 Aerospace Stocks.}
    \label{fig:techIndicator}
\end{figure}

Figure \ref{fig:company_names} shows a visualization of the results from context-based sentiment analysis on non-aerospace company names for Boeing financial text documents. We can see that the relative sentiment values for companies like Korea Electric Power Corp and Banco Santander SA are significantly higher than other company names mentioned in financial news regarding Boeing stock. Intuitively, these results show that when financial news articles include these company names and Boeing stock together, they are mentioned in a positive view and often lead to financial gain for Boeing.

We perform data pre-processing and compute various statistical features from our raw numerical stock data in order to generate additional technical indicators for our model. For example, we calculate the 7-day and 21-day moving averages of all stocks for the timeframe we selected. We also calculate the MACD or Moving Average Convergence Divergence by subtracting the 26-period Exponential Moving Average (EMA) from the 12-period EMA. The general formula for EMA is \begin{equation*}
  \text{EMA} = \text{Price}(t_n) \times k + \text{EMA}(t_{n-1}) \times (1-k),
\end{equation*}
where $t_n$ and $t_{n-1}$ denotes the $n^{th}$ and ${n-1}^{th}$  timestep, $k$ is defined as $2 / (N + 1)$ and $N$ denotes the number of timesteps in the EMA period. We also calculate the upper and lower Bollinger bands by adding and subtracting the 20-day rolling standard deviation from the 21-day moving average, respectively. A visualization of the technical indicators for the eight aerospace stocks over the dataset time period is shown in Figure \ref{fig:techIndicator}. 

We extract global and local trends in our stock prices using Fourier transforms, which take a function and decompose it into various sine waves that approximate the function. Fourier transforms are often used as a financial forecasting tool, which convert time-series data to frequencies and amplitudes of patterns. We decompose the historical close prices of the eight stocks into various sine functions of various amplitudes and frames. The Fourier transform of a function $g(t)$ is defined as 
\begin{equation*}
    G(\omega) = \int^{\infty}_{-\infty} g(t)e^{-i2\pi \omega t} dt.
\end{equation*}

\begin{figure}
    \centering
    \includegraphics[width=\textwidth,trim={2cm 2cm 2cm 2cm},clip]{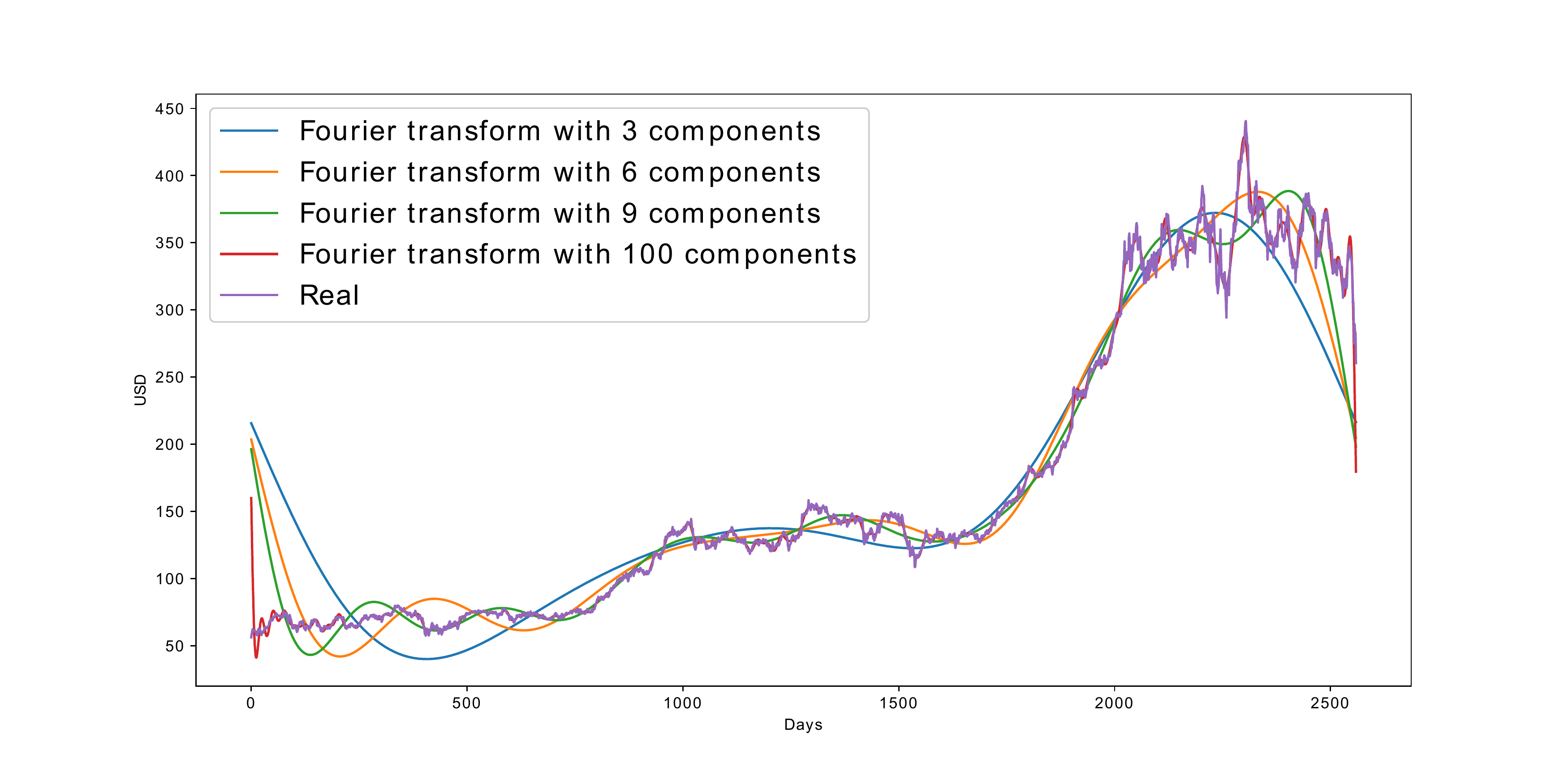}
    \caption{Various Fourier Transform Sine Wave Decompositions for Historical Daily Boeing Stock Close Price.}
    \label{fig:fourier}
\end{figure}

Figure \ref{fig:fourier} displays various Fourier transform sine wave function decompositions for historical Boeing stock close prices. 

We also gathered additional numerical features by running an Auto Regressive Integrated Moving Average (ARIMA) model to forecast future stock prices using the historical stock price data. Although the forecasted stock prices were not exceedingly accurate, we are able to include the auto-regressive parameters for the best fit ARIMA model into our deep learning model. We find that an ARIMA(5,1,0), or an ARIMA model with 5 autoregressive terms, first order differencing, and no moving averages, performs the best with the given data. We include information on the autocorrelation and partial autocorrelation functions used to determine the optimal auto-regressive parameters into our feature set as well. 

\subsection{Implementation}\label{3two}

We create a Naive Bayes' classifier to perform Naive Bayes' sentiment analysis
on the vectorized representations of the financial text documents. We utilize
one-hot encoding to vectorize each sentence of the text dataset in order to
predict sentiments on a sentence-by-sentence level. The loss function we used to
optimize the most likely sentiment, or class, is defined in \eqref{eq:1}. 

\begin{figure}
    \centering
    \includegraphics[width=0.9\textwidth]{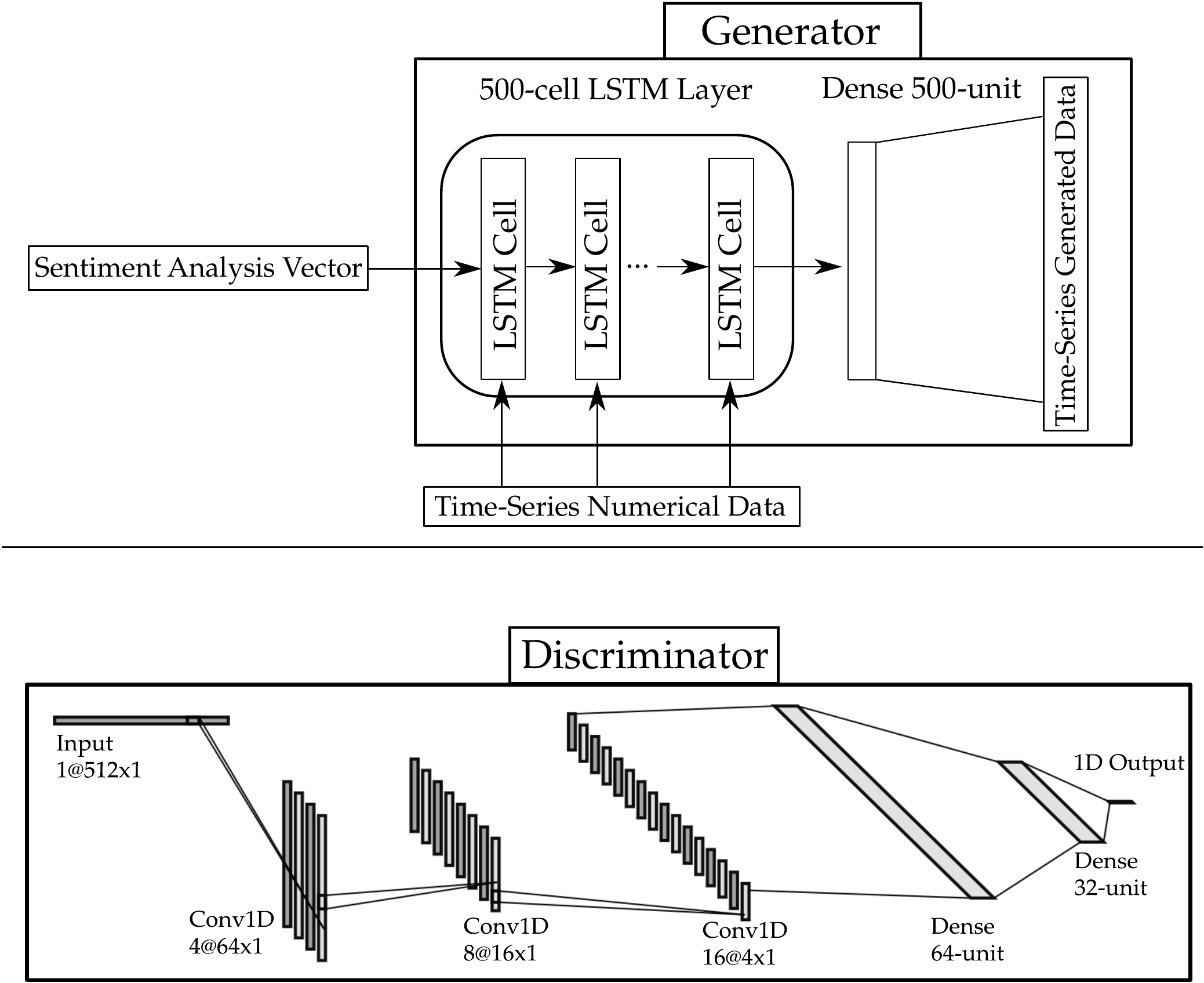}
    \caption{Model Architecture of the LSTM Generator Network and CNN Discriminator Network of our GAN Model.}
    \label{fig:modelArch}
\end{figure}

For the second component of our model, we used a Long-Short Term Memory (LSTM) for the Generator network and a Convolutional Neural Network (CNN) for the Discriminator network. The Generator architecture consists of an LSTM layer with 128 input units for the 128 daily numerical features processed from historical data for the eight aerospace stocks and a Dense layer with 1 output to generate the stock price for a given time. The Discriminator architecture consists of a 1-dimensional CNN with three Convolution layers and two Dense layers with one output to generate the classification signal. The structure of the GAN model is illustrated in Figure \ref{fig:modelArch}. The hyperparameter values for our model after tuning is shown in Table \ref{tab:hyperparameter}. 

\begin{table}
    \centering
    \small
    \begin{tabular}{l c c}
    \hline
        \bf{Hyperparameter} \hspace{5cm} & \bf{Value}  \\
    \hline
        LSTM Weight Initializer \hspace{5cm} & Xavier \\
        LSTM Hidden Units \hspace{5cm} & 500 \\
        LSTM Sequence Length (in days) \hspace{5cm} & 30 \\
        LSTM Learning Rate \hspace{5cm} & 0.01 \\
        LSTM Optimizer \hspace{5cm} & Adam \\
        LSTM Regularization \hspace{5cm} & L1 (Lasso) \\ 
        Conv1D Kernel Size \hspace{5cm} & 5 \\
        Conv1D Stride \hspace{5cm} & 2 \\
        Conv1D Padding \hspace{5cm} & None \\
        Conv1D Activation Function \hspace{5cm} & LeakyReLU ($\alpha = 0.01$) \\
        Dense Activation Function \hspace{5cm} & ReLU \\
        BatchNorm Momentum \hspace{5cm} & 0.9 \\
        BatchNorm Epsilon \hspace{5cm} & $1 \times 10^{-5}$ \\
        Epochs \hspace{5cm} & 500 \\
        Batch Size \hspace{5cm} & 16 \\
    \hline
    \end{tabular}
    \caption{Model Hyperparameter Values}
    \label{tab:hyperparameter}
\end{table}

\subsection{Results}\label{3three}

We seek to predict future stock prices for a variety of time frames using both financial news texts and numerical features. We present results and error margins for predicting future stock prices 1 day in the future, 15 days in the future, and 30 days in the future. It is important to note that no additional ground truth data is given within the 15 or 30 day prediction time frame, so the model uses its own previous days' predictions. By the nature of Long Short-Term Memory (LSTM) models, our model provides predictions on a daily time frequency, and this measure ensures that we are not performing data leakage when predicting for 15 and 30 day forecast horizons. Our results show that our model has the lowest average error and thus is the best performing model among all baselines and existing deep learning models tested. To measure the accuracy of our model, we used the Root Mean Square Error (RMSE) and Normalized RMSE (NRMSE) error metrics. RMSE error is calculated as 
\begin{equation*}
  \text{RMSE} = \sqrt{\sum_{i=1}^{n} \frac{(\hat{y}_i - y_i)^2}{n}},
\end{equation*}
where $n$ is the number of observations, $\hat{y}$ is the predicted value, and $y$ is the ground truth. NRMSE is calculated as 
\begin{equation*}
  \text{NRMSE} = \frac{\text{RMSE}}{\bar{y}},
\end{equation*}
where $\bar{y}$ is the mean of the ground truth.

We included the NRMSE error metric in our error analysis because it allows us to compare our model accuracies against other models of different scales and prediction targets. Table \ref{tab:results} shows the RMSE and NRMSE results of our model on various N-day forecast horizons where $N \in \{1,15,30\}.$ 
\begin{table}
    \small
    \centering
    \begin{tabular}{l l  r  r  r }
    \hline
      Metric \hspace{3cm}  &  Model  & N = 1 & N = 15 & N = 30 \\
    \hline 
              &   \bf{ST-GAN} & $\bf{0.16}$ & $\bf{2.39}$ & $\bf{4.37}$ \\
      & GAN  & 0.74 & 11.74 & 20.41 \\
        & FC-LSTM  & 0.41 & 6.13 & 13.24 \\
      RMSE & ARIMA(5,1,0)  & 1.94 & 19.34 & 32.43 \\
       & Sentiment Analysis & 6.89 & 90.24 & 174.87 \\
       & GAN-FD & 0.28 & 3.41 & 8.25 \\
       & VolTAGE & 0.33 & 7.25 & 5.11 \\
       & DP-LSTM & 0.65 & 5.34 & 14.09 \\
       \hline
                &   \bf{ST-GAN}     &   $\bf{0.00049}$ & $\bf{0.00751}$ & $\bf{0.01326}$ \\
      & GAN &  0.00229 & 0.03693 & 0.06193 \\
        & FC-LSTM & 0.00127 & 0.01928 & 0.04018 \\
      NRMSE & ARIMA(5,1,0) & 0.00600 & 0.06083 & 0.09841 \\
       & Sentiment Analysis & 0.02133 & 0.28383 & 0.53063 \\
       & GAN-FD & 0.00196 & 0.01114 & 0.02961 \\
       & VolTAGE & 0.00152 & 0.04241 & 0.01441 \\ 
       & DP-LSTM & 0.00162 & 0.00993 & 0.05114 \\
       \hline
    \end{tabular}
    \caption{RMSE and NRMSE Error Values for $N$-day Forecast Horizons on Boeing Stock}
    \label{tab:results}
\end{table}

We show various baseline model error results against our model, \emph{ST-GAN} or
Stochastic Time-series Generative Adversarial Network. The baseline models were
all ran in the same experimental setting as described for our model. In addition
to the generic baselines (GAN, FC-LSTM, ARIMA, and Sentiment Analysis), we also
show results of running other existing models in stock price forecasting in our experimental environment. We show RMSE and NRMSE error values for the GAN-FD model proposed by \citet{zhou2018stock} that uses a GAN model with an LSTM Generator network and a CNN Discriminator network on numerical stock price data. We also tested the VolTAGE model, proposed by \citet{sawhney2020voltage}, that uses a Graph Convolutional Network (GCN) model on Earnings Conference Call data and historical stock prices in our experimental setting. Finally, we tested the DP-LSTM model proposed by \citet{li2019dp} that uses a modified LSTM on time-series historical stock price and financial news data. 

\begin{figure}
    \centering
    \setlength\abovecaptionskip{-5pt}
    \includegraphics[width=0.9\textwidth]{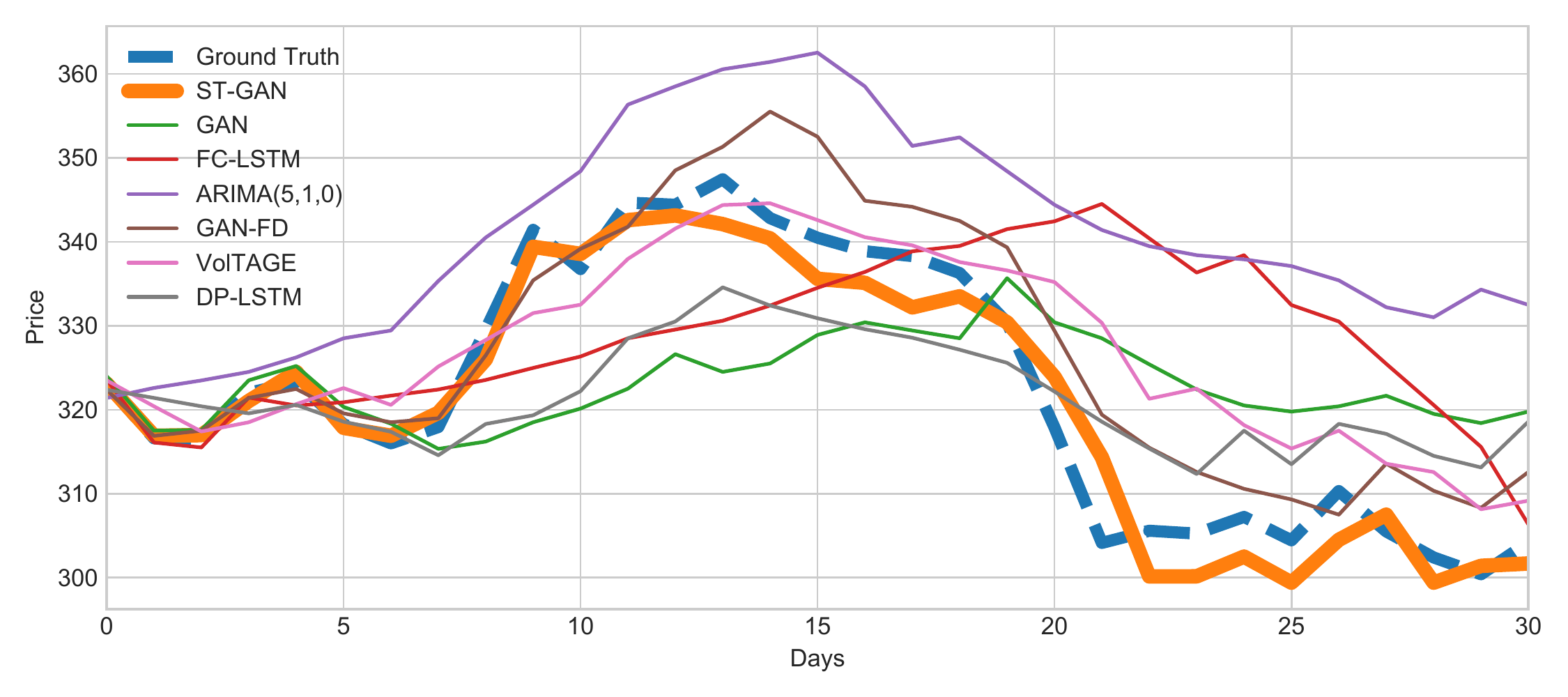}
    \caption{ST-GAN Performance Against All Experimental Models 30-day Forecast Horizon Predicted Boeing Stock Price Graph (1/24/20 - 2/25/20)}
    \label{fig:allModel}
    \centering
    \includegraphics[width=0.9\textwidth]{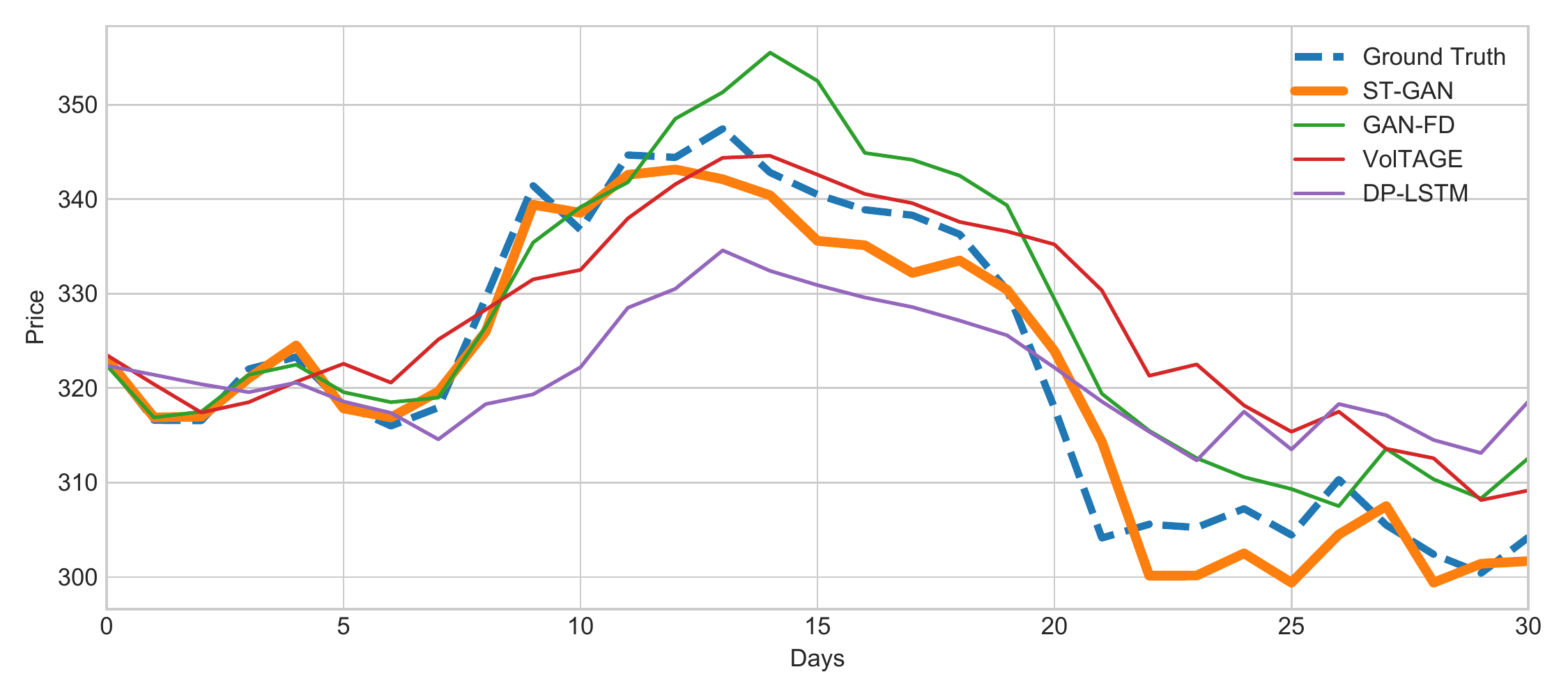}
    \caption{ST-GAN Performance Against Previous Research Models 30-day Forecast Horizon Predicted Boeing Stock Price Graph (1/24/20 - 2/25/20)}
    \label{fig:prevModel}
\end{figure}

Figure \ref{fig:allModel} shows a graph of the predicted stock prices for Boeing stock in a 30-day forecast horizon from January 25, 2020 to February 24, 2020. We show the performance of the baseline models against our model, \emph{ST-GAN}.

We also show the performance of our model against previous research models only in Figure \ref{fig:prevModel}. 

\section{Conclusion}\label{four}

In this work, we have introduced a novel deep learning model that analyzes financial news texts and financial numerical data to predict future aerospace stock trends in the short and long term with unparalleled accuracies. We proposed a novel architecture that applies Naive Bayes' sentiment analysis to financial news texts and uses the learned representations alongside financial numerical data to train a Generative Adversarial Network (GAN) for time-series prediction of stock prices. Our experimental results show that our model, which utilizes cutting-edge technologies, may have a significant impact on the practice of portfolio management. 

Our error analysis using the RMSE and NRMSE error metrics shows significant improvement over existing deep learning models in this field using similar technologies. We believe our contributions can be applied to help researchers, scholars, portfolio managers, investment officers, trustees, and consultants make better decisions and predictions in the financial sector. 

In the future, we expect to measure the isolated impact of our model's robust textual understanding from sentiment analysis on the overall prediction accuracy. We also hope to increase the temporal frequency of our predictions and expand our model prediction targets to a more diverse portfolio of stocks.

\bibliographystyle{apalike}
\bibliography{ref}

\end{document}